# Magnons and Phonons Optically Driven Out of Local Equilibrium in a Magnetic Insulator

Kyongmo An[1], Kevin S. Olsson[1], Annie Weathers[2], Sean Sullivan[3], Xi Chen[3], Xiang Li[3], Luke G. Marshall[3*], Xin Ma[1], Nikita Klimovich[1], Jianshi Zhou[2,3], Li Shi[2,3,†], and Xiaoqin Li[1,3,†]

[1]*Department of Physics, Center for Complex Quantum Systems, The University of Texas at Austin, Austin, Texas 78712, USA*
[2]*Department of Mechanical Engineering, The University of Texas at Austin, Austin, Texas 78712, USA*
[3]*Materials Science and Engineering Program, Texas Materials Institute, The University of Texas at Austin, Austin, Texas 78712, USA*

The coupling and possible non-equilibrium between magnons and other energy carriers have been used to explain several recently discovered thermally driven spin transport and energy conversion phenomena. Here, we report experiments in which local non-equilibrium between magnons and phonons in a single crystalline bulk magnetic insulator, $Y_3Fe_5O_{12}$, has been created optically within a focused laser spot and probed directly via micro-Brillouin light scattering. Through analyzing the deviation of the magnon number density from the local equilibrium value, we obtain the diffusion length of thermal magnons. By explicitly establishing and observing local non-equilibrium between magnons and phonons, our studies represent an important step toward a quantitative understanding of various spin-heat coupling phenomena.

The emerging field of spin caloritronics has been stimulated by a number of recent discoveries, such as large magnon thermal conductivity [1,2], spin Seebeck effect (SSE) [3-10], spin Peltier effect [11,12], magneto-Seebeck effect [13], and thermal spin transfer torque (STT) [14,15]. These discoveries hold promise for new technologies based on thermally driven spin transport phenomenon. For example, thermal-STT improves upon current STT-based memory devices by reducing the threshold current for STT-induced magnetic switching [16]. In addition, the large magnon thermal conductivity observed in some cuprate crystals can find potential applications for thermal management [1]. Moreover, the spin Seebeck and spin Peltier effects are being explored for applications in novel thermoelectric energy conversion devices.

The coupling and non-equilibrium between different energy carriers, namely magnons, phonons, and electrons plays an important role in the current theories of spin caloritronic phenomena. For example, in the magnon-mediated transverse SSE model [4,17], it is speculated that the magnon population is out of local equilibrium with the phonon bath. However, such local non-equilibrium has not been directly observed [18]. Indeed, to drive and observe non-equilibrium between magnons and phonons, a temperature gradient must be established within a length scale smaller than the distance over which magnons relax toward complete thermodynamic equilibrium with the phonon bath. Experimental studies on creating and detecting local magnon-phonon non-equilibrium, and probing its associated fundamental length scale, can help to further advance the field of spin caloritronics.

In this letter, we demonstrate that magnons and phonons can be driven out of local equilibrium in a bulk crystal of the magnetic insulator $Y_3Fe_5O_{12}$ (yttrium iron garnet, or YIG) that is irradiated by a focused laser beam to obtain a temperature gradient on the order of $10^6$ K m$^{-1}$, two orders of magnitude larger than those achieved in previous experiments [18,19]. Using the Brillouin light scattering (BLS) technique, we are able to directly probe both the phonon temperature and magnon number density at the same location. Our measurements show that the magnon number density in the laser spot is apparently lower than the local equilibrium value. The fact that one can drive magnons and phonons out of local equilibrium by localized heating within a few microns suggests that the characteristic coupling length between thermal magnons and phonons is comparable to or longer than a few microns in YIG. The measured deviation in magnon number density from equilibrium also allows us to obtain a thermal magnon diffusion length of about 3 μm based on a diffusion model. These findings are essential for reaching a microscopic understanding of a host of spin caloritronic phenomena and exploring their potential device applications.

We measured the temperature dependent magnon and phonon spectra using the micro-BLS technique as shown in Fig. 1 [19-21]. A green laser with a wavelength of $\lambda = 532$ nm and power of 8 mW was used as the probing laser in all BLS measurements. An additional red laser with $\lambda = 660$ nm was used to create local heating in some of the measurements. The power of the heating laser was varied from 0 to 19.1 mW. The beam sizes for the probing and heating lasers were obtained by fitting the beam intensity with a Gaussian function and were found to have an effective radius of $w_g = 0.8$ μm and $w_r = 1.3$ μm, respectively. The YIG sample was oriented with the [110] direction normal to the surface, and an external magnetic field of 49.5 mT was applied along the [1$\bar{1}$0] in-plane direction of the sample in all measurements.

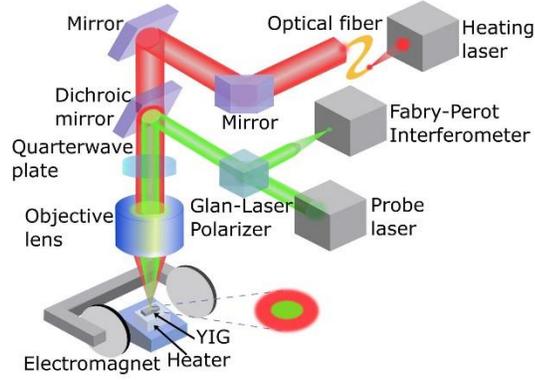

FIG. 1 Schematic of micro-BLS setup for measuring magnon and phonon BLS spectra in YIG under stage heating or focused laser heating.

We first discuss the magnon and phonon modes probed in the BLS spectra shown in Fig. 2. The dominant phonon and magnon modes probed by BLS have the same wave vector ($q$) determined by the change in the photon momentum, as required by momentum conservation, $q = k_s - k_i$, where $k_s$ and $k_i$ are the scattered and incident wave vectors of the laser averaged over the light cone, respectively. In our backscattering geometry, $k_s$ and $k_i$ are nearly anti-parallel, thus $q$ equals $4\pi n/\lambda = 5.53 \times 10^7$ m$^{-1}$, where $n = 2.34$ is the index of refraction for YIG at $\lambda = 532$ nm [22]. This wave vector translates to a wavelength of 113 nm for both the magnon and phonon modes probed. The calculated frequencies of the probed magnon and phonon modes agrees well with our experimental observations (see the Supplemental Material [23] for details).

We first uniformly heated the YIG sample with an external heating stage while recording the temperature dependent BLS spectra, as shown in Fig. 2. While the phonon and magnon populations increase with temperature according to the Bose-Einstein distribution, the intensity, linewidth, and frequency of the BLS peak show complicated temperature dependence [24,25]. In Fig. 2a, the peak frequency in the magnon BLS spectra shifts downwards in frequency by 0.25 GHz when the temperature is increased from 302 K to 345 K. This shift arises from the reduction in the saturation magnetization with increasing temperature. As each magnon reduces the magnetic moment by the same amount, the measured magnon frequency shift can be used to probe the change in the local magnon number density. We note, however, that probing the magnon number density is insufficient to determine the magnon temperature when local non-equilibrium exists between magnons and the lattice within a length scale shorter than the magnon diffusion length. Within this length scale, a non-zero magnon chemical potential can produce an approximately constant magnon number density as the magnon temperature is changed [26,27].

The phonon BLS spectra exhibit a downward frequency shift of 0.14 GHz as the temperature increases from 302 K to 345 K as shown in Fig. 2b. Because the phonon frequency shift is caused by bond softening and anharmonic coupling among phonon modes, the phonon frequency shift is influenced by the occupancy of all other phonon modes that are coupled to the long-wavelength phonon mode probed by BLS [24]. As such, the phonon peak shift can be used to probe the average temperature of the broad spectrum of thermally excited phonons instead of only the temperature corresponding to the long-wavelength mode directly probed by BLS.

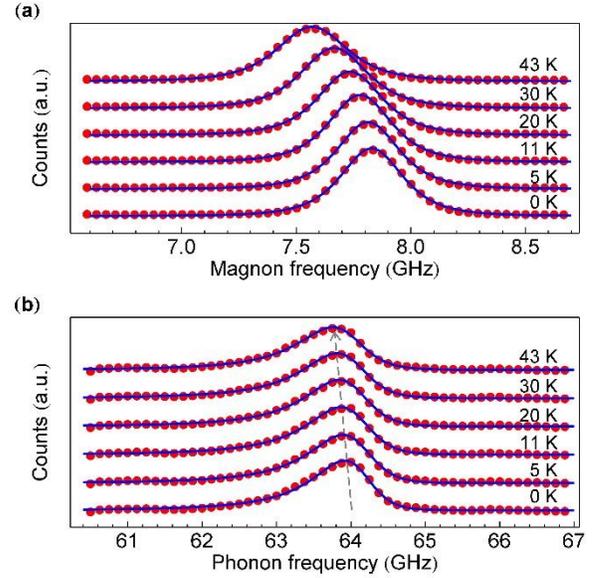

FIG. 2 Temperature dependent BLS spectra for (a) magnons and (b) phonons, obtained with the YIG sample heated uniformly on a heater stage. Solid lines are fitting using (a) symmetric and (b) asymmetric squared Lorentzian functions (see the Supplemental Material [23]). The arrow in (b) is drawn to show the small downward frequency shift of the phonon signal. The numbers inside the figure represent the stage temperature rise from the room temperature.

In order to create and probe local non-equilibrium between magnons and phonons, we measured the phonon and magnon BLS spectra with the addition of a red heating laser while maintaining the sample stage at room temperature. Both the magnon and phonon frequencies shift down with increasing heating laser power (Fig. 3b) similar to the stage heating condition (Fig. 3a). Figure 3c shows the equivalent stage temperature rise for the stage heating condition that yields the same phonon frequency shift as that in the red laser heating experiment at each laser power. The difference in strain between the stage and laser heating cases were taken into account in Fig. 3c-d. (see the Supplemental Material [23] for detailed calculations of the strain effects and BLS measurements under hydrostatic pressure)

We now compare magnon frequency shifts in the two heating configurations. Figure 3d shows the magnon frequency shift as a function of the strain-corrected phonon frequency shift based on Fig. 3a and 3b. Because the measured phonon frequency shift is determined by

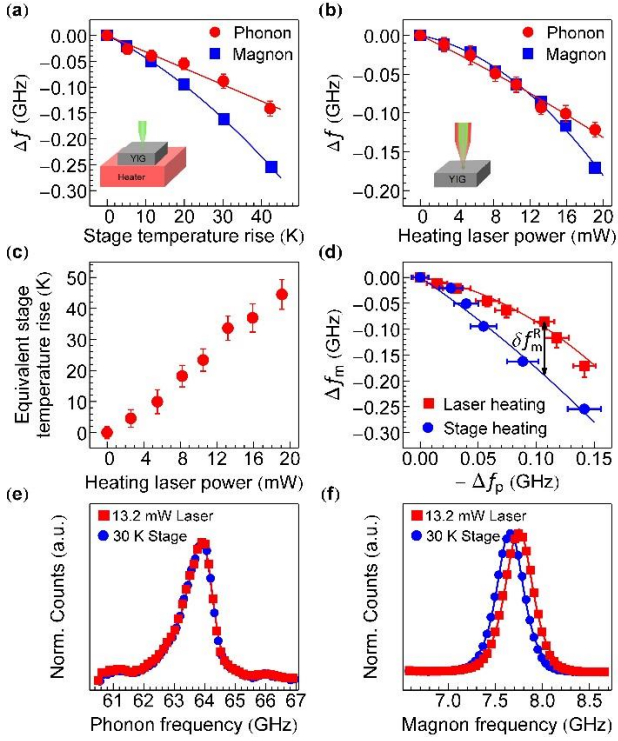

FIG. 3 (a) Peak frequency changes of magnons and phonons as a function of the sample stage temperature rise. The red line is a linear fitting of the phonon data while the blue line is a quadratic fitting of the magnon data. (b) Peak frequency changes for magnons and phonons as a function of the red heating laser power when the stage is kept at room temperature. The lines are polynomial fitting of the data. (c) Equivalent stage temperature rise in the stage heating measurement yielding the same strain-corrected phonon frequency downshifts in the laser heating measurement is plotted as a function of the heating laser power. (d) Magnon peak frequency changes ($\Delta f_m$) as a function of phonon peak frequency changes ($\Delta f_p$) in the two different heating configurations after the strain effects are accounted for. The arrow shows the difference ($\delta f_m^R > 0$) in the magnon frequency between the 13.2 mW red laser heating and the corresponding stage heating that yields the same strain-corrected $\Delta f_p$. Raw BLS spectra of (e) phonon and (f) magnon under 30 K stage temperature rise (blue) and 13.2 mW laser heating (red). Solid lines are fitting. The spectra are normalized to their maximum counts.

the local phonon temperature rise when the strain effect is accounted for, the same $x$-axis value in Fig. 3d represents the same average local phonon temperature rise in the probe laser spot in two different heating experiments. It is apparent that the magnon frequency shift is smaller for the laser heating case than for the uniform stage heating condition. Specifically, we show two phonon/magnon spectra that have very similar averaged temperatures under stage heating (30 K) and laser heating (13.2 mW) in Fig. 3e and 3f. While the phonon spectral shapes are nearly identical (Fig. 3e), the magnon spectra are clearly shifted (Fig. 3f). As the magnon frequency shift is directly related to the magnon number density, this result clearly shows that the local magnon number density within the heating laser spot is not relaxed to the equilibrium value at the local phonon temperature. In comparison, two prior BLS measurements have been conducted in an attempt to directly verify the existence of local non-equilibrium in either magnetic insulators [18] or metals [19]. However, the ~$10^4$ K m$^{-1}$ temperature gradients applied over mm scale were not sufficient to produce observable non-equilibrium between magnons and phonons.

To describe the established non-equilibrium between magnons and phonons quantitatively, we adopt the following steady state magnon diffusion equation, after using an averaged diffusion length $l_r$ for the broad spectrum of thermal magnons [28],

$$\nabla^2 n = \frac{n - n_0}{l_r^2}, \quad (1)$$

where $n$ is the magnon number density, and $n_0$ is the magnon number density in local equilibrium with the phonon temperature $T_p(\boldsymbol{r})$, which is calculated using the Bose-Einstein distribution $n_0 = \sum_{\boldsymbol{k}}\left[\exp\left(\frac{\hbar\omega(\boldsymbol{k})}{k_B T_p(\boldsymbol{r})}\right) - 1\right]^{-1}$, where $\hbar$ and $k_B$ are the reduced Planck's constant and Boltzmann constant, respectively, $\hbar\omega(\boldsymbol{k})$ is the spectroscopic energy of the magnon mode with wave vector $\boldsymbol{k}$, and the summation is over all magnon modes. Each magnon decreases the magnetic moment by $g\mu_B$, where $g$ and $\mu_B$ are the Landé $g$-factor and the Bohr magneton, respectively. Therefore, the deviation from local equilibrium, $\delta n \equiv n - n_0$, leads to a deviation in the local magnetization, $\delta M = -g\mu_B \delta n$. This deviation further gives rise to a deviation in the magnon peak frequency, $\delta f_m \approx \left(\frac{\partial f_m}{\partial M}\right)_H \delta M = -\left(\frac{\partial f_m}{\partial M}\right)_H g\mu_B \delta n$. Under the experimental conditions, we find that $\nabla^2 n_0 \approx \left(\frac{\partial n_0}{\partial T_p}\right)_H \nabla^2 T_p$ is an accurate approximation [23]. Hence, Eq. (1) can be rewritten as

$$\nabla^2 T_p + \nabla^2 \delta\theta_m = \frac{\delta\theta_m}{l_r^2},$$

where $$\delta\theta_m \equiv \frac{\delta f_m}{\left(\frac{\partial f_m}{\partial T_p}\right)_H}. \quad (2)$$

Here, $\left(\frac{\partial f_m}{\partial T_p}\right)_H$ depends on the local phonon temperature, $T_p(r, z)$, and was obtained from the measurements of $f_m$ at different stage temperatures with the heating from the green probe laser accounted for [23]. Equation (2) can be used to solve for $\delta\theta_m(r, z)$, and subsequently $\delta f_m(r, z)$.

The phonon temperature profile is determined by the steady state energy equation,

$$\nabla \cdot (\kappa_p \nabla T_p) + \nabla \cdot (\kappa_m \nabla T_m) + Q = 0, \quad (3)$$

where $\kappa_m$ and $\kappa_p$ are the magnon and phonon contributions to the total thermal conductivity $\kappa = \kappa_p + \kappa_m$, and $Q$ is the power density of the absorbed laser. Here, $T_m$ is the magnon temperature. It has previously been found that $\kappa_m$ is much smaller than $\kappa_p$ in YIG near room temperature [29,30]. In addition, $|\nabla T_p|$ is greater than $|\nabla (T_p - T_m)|$. Hence, the energy equation can be approximated as (see the Supplemental Material [23] for details of the energy equation),

$$\nabla \cdot (\kappa \nabla T_p) + Q \approx 0. \quad (4)$$

Eq. (2) and (4) were solved numerically using the commercial software COMSOL for both the red laser heating experiment at 13.2 mW and the stage heating condition with 33 K stage temperature rise, both of which yield the same strain-corrected phonon frequency shift in the probe laser spot according to Fig. 3c. The calculations were based on independently measured temperature-dependent thermal conductivity and absorption coefficients, and the green probe laser present in both heating configurations was taken into account explicitly. Figure 4a shows the difference in the local phonon temperature rise in the two simulations, $\Delta T_p^R(r,z)$, caused by the red heating laser. The weighted average value in the probe laser spot, $\langle \Delta T_p^R(r,z) \rangle$, is 32 K, which agrees with the equivalent stage temperature rise of 33 ± 4 K shown in Fig. 3c for the experiment at 13.2 mW heating laser power. The agreement verifies the phonon temperature measured by the BLS method. The corresponding magnon frequency deviation profile between the two heating conditions, $\delta f_m^R(r,z)$, is shown in Fig. 4b for $l_r = 3.1\ \mu m$. The weighted average value $\langle \delta f_m^R(r,z) \rangle$ is plotted as a function of $l_r$ in Fig. 4c. The black solid line and gray shaded area are shown to indicate the mean of the measured $\delta f_m^R$ and its uncertainty, respectively. From the intersection, we find $l_r = 3.1 \pm 0.9\ \mu m$. The diffusion length obtained in our measurements corresponds to the value at about 372 K, due to the heating by both lasers. Because of the relatively long spin diffusion length compared to the heating laser spot size, the calculated magnon number density in the red laser heating experiment is lower than that in the stage heating condition, as shown in Fig. 4d, despite the same average phonon temperature in the probe laser spot. Details of the numerical calculations can be found in the Supplemental Material [23].

Our finding has important implications for spin caloritronics, where various relaxation processes of magnons play an important role. The scattering of magnons occurs by both spin-conserving and non-spin-conserving processes. Spin-conserving processes only relax energy and momentum. Such processes occur frequently and exhibit a length scale of a few nanometers according to thermal conductivity measurements [29]. On the other hand, non-spin-conserving processes relax the magnon number density through spin-orbit interactions with the lattice bath and may exhibit a longer length scale [31,32]. A long magnon spin diffusion length is considered essential for SSE [7,17]. While it has been suggested that this length could be as long as millimeters based on the initial SSE measurements [3,9], a recent measurement of magnon transport in a transverse SSE geometry has reported a magnon diffusion length of ~8 $\mu m$ for thermally excited magnons in YIG near room temperature [33]. However, it has been suggested that the interpretation of transverse SSE measurements can be complicated by the presence of unwanted longitudinal thermal gradients in addition to the intended transverse gradient in the experimental setup [34].

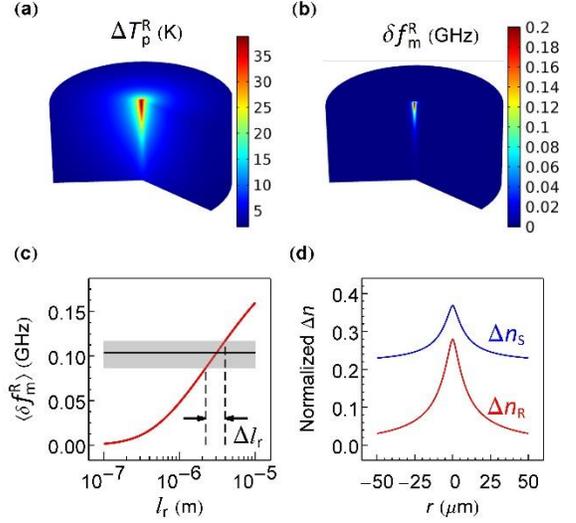

FIG. 4 Simulated spatial profiles of (a) the difference in the local phonon temperature rise in two simulations caused by the red laser heating and (b) magnon frequency deviation between the two heating conditions calculated with $l_r = 3.1\ \mu m$. The displayed volume is 50 $\mu m$ in both the radial and axial directions in both spatial profiles. (c) Weighted averages of $\delta f_m^R$ over the probe laser spot as a function of $l_r$. The red line shows the relation between each $l_r$ used in the simulations and the calculated $\langle \delta f_m^R \rangle$. The gray area represents the uncertainty of measured $\delta f_m^R$ while the black solid line is the mean of measurements. $\Delta l_r$ is the possible range of $l_r$ for the calculated values due to the uncertainty of the measurement. (d) Simulated profiles of the magnon population rise relative to the room temperature value at the surface of YIG. The red and blue lines represent the magnon population rises for the red laser heating experiment ($\Delta n_R$) and the stage heating condition ($\Delta n_S$), respectively, as a function of position including the effect of the green probe laser. The magnon population rises are normalized by the room temperature value.

In summary, our experiments clearly demonstrate that magnons can be driven out of local equilibrium with the phonon bath in YIG when a large temperature gradient is generated on the scale of a few microns. Furthermore, our analysis shows that the observed local non-equilibrium is associated with a long thermal magnon spin diffusion length on the order of several microns. The optically-based non-contact method for exciting and probing non-equilibrium magnon transport demonstrated here can be extended to various materials where such properties remain unknown.

The authors thank David G. Cahill and Yaroslav Tserkovnyak for their insightful discussions on the strain effects and magnon diffusion theory, and acknowledge Jung-Fu Lin for valuable advice on the pressure cell design. KA, AW, XC, JZ, and LS were supported by the Army Research Office (ARO) MURI program under Award # W911NF-14-1-0016. Part of the support for KA and support for XM and XQL was provided as part of the SHINES, an Energy Frontier Research Center funded by the U.S. Department of Energy (DOE), Office of Science, Basic Energy Science (BES) under Award # DE-SC0012670. Work by KSO and SS was supported by National Science Foundation Thermal Transport Program under Award # CBET-1336968. AW is supported by an NSF Graduate Research Fellowship. The BLS instrumentation for the measurements under different heating was obtained via support from Airforce Office of Scientific Research (AFOSR) under grant # FA9550-08-1-0463 and FA9550-10-1-0022. The BLS instrument for the pressure-dependent measurement was obtained via support from ARO grant # W911NF-14-1-0536.

*Present address: Department of Chemical Engineering, Northeastern University, Boston, Massachusetts 02115, USA
†Authors to whom correspondence should be addressed. Email: elaineli@physics.utexas.edu, lishi@mail.utexas.edu

# Supplemental Material: Magnons and Phonons Optically Driven Out of Local Equilibrium in a Magnetic Insulator

Kyongmo An[1], Kevin S. Olsson[1], Annie Weathers[2], Sean Sullivan[3], Xi Chen[3], Xiang Li[3], Luke G. Marshall[3*], Xin Ma[1], Nikita Klimovich[1], Jianshi Zhou[2,3], Li Shi[2,3], and Xiaoqin Li[1,3]

[1]*Department of Physics, Center for Complex Quantum Systems, The University of Texas at Austin, Austin, Texas 78712, USA*
[2]*Department of Mechanical Engineering, The University of Texas at Austin, Austin, Texas 78712, USA*
[3]*Materials Science and Engineering Program, Texas Materials Institute, The University of Texas at Austin, Austin, Texas 78712, USA*
*current address: Department of Chemical Engineering, Northeastern University, Boston, Massachusetts 02115, USA*

## Table of Contents

**S1 Sample preparation**

Single crystals of YIG were grown via the traveling solvent floating zone method by using an infrared-heated image furnace [S1]. The use of a solvent was necessary since YIG exhibits peritectic melting. The YIG feed and seed rods were prepared by the solid-state reaction of $Fe_2O_3$ and $Y_2O_3$ powders at 1400 ˚C for 24 hours in oxygen, while the solvent (3 $Fe_2O_3$ : 17 $Y_2O_3$) was calcined at 1100 ˚C for 24 hours in oxygen. Following calcination, the powders were carefully ground and hydrostatically pressed into rods under 20 MPa of pressure, which were then sintered at 1400 ˚C and 1100 ˚C in oxygen for the feed/seed rods and the solvent rods, respectively. Growth was then carried out under an $O_2$ flow of ~1 standard cubic foot per hour (SCFH) with a growth rate of 0.95 mm/hr. Powder X-ray diffraction (XRD) was performed on a powder sample obtained by crushing a small piece of the as-grown ingot. All the diffraction peaks can be indexed to the garnet structure $Y_3Fe_5O_{12}$ with space group *Ia-3d*. The refinement of the XRD pattern gives the lattice parameter *a* = 12.370 Å, which is consistent with the value in the literature [S2]. Laue back reflection was used to check the crystal quality and orient the crystal. An oriented chip of YIG crystal with dimensions 4 mm × 2 mm × 0.4 mm was cut and polished with the [110] axis normal to the polished surface.

**S2 Thermal conductivity measurement**

The thermal conductivity of the YIG sample was measured in the temperature interval between 100 K and 300 K by a steady-state method. The sample size was approximately 0.5 × 0.5 × 2.5 $mm^3$. The reference was a rod of constantan alloy with a diameter of 0.5 mm. The differential thermocouple was made of copper and constantan wires. The measured thermal conductivity of the YIG single-crystal along the [111] direction is 8.8 ± 1.3 W $m^{-1}$ $K^{-1}$ at 300 K. For the thermal conductivity above room temperature, we took the thermal conductivity data of YIG along [111] from a previous work [S3]. These results are plotted together with our measured data in Fig. S1. By fitting the data for temperatures above 200 K, we obtained the high temperature thermal conductivity as a function of temperature, $\kappa(T) = a + b/T$, where $a = 0.802 \text{ W m}^{-1}\text{K}^{-1}$ and $b = 2385 \text{ W m}^{-1}$, as shown as the solid line in Fig. S1. This temperature dependent thermal conductivity was then used in our numerical simulations.

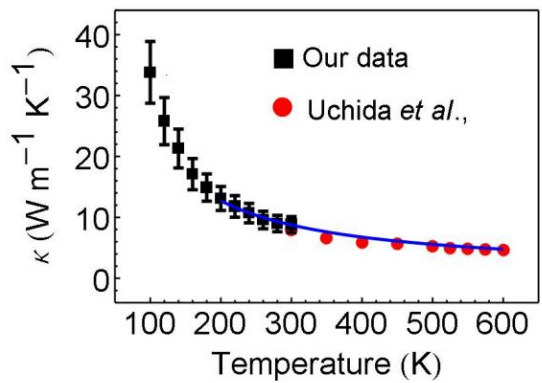

Fig. S1 Thermal conductivity of the YIG single crystal along [111] direction. The solid line is a fit to the high temperature data.

**S3 Micro-BLS measurements**

Only the anti-Stokes side of the spectra was collected, because the Stokes side showed the same modes with only small variance in intensity. The third free spectral range of a tandem Fabry-Pérot interferometer was used to improve the frequency resolution in all BLS measurements. The single crystal YIG sample was tilted by 30 degrees from normal incidence to reduce the background signal arising from elastically scattered light. The polarizations of the photons scattered by magnons and phonons are orthogonal [S4], thus both spectra can be acquired by placing a quarter wave plate in the collection path. The quarter wave plate was kept at an angle of about 30 degrees between its slow axis and the incident polarization of light, so as to take both phonon and magnon spectra without changing the incident laser power, which can be attenuated by the angle of the quarter wave plate. Magnon and phonon spectra were taken alternately by changing the frequency window in the Fabry-Pérot interferometer.

**S4 Elliptical laser spot on YIG surface**

Because the sample was tilted from the direction of normal incidence to reduce background noise introduced by elastically scattered light, the laser spot was elliptical rather than circular as assumed in the calculation. The beam sizes for the probing and heating lasers were obtained using a Gaussian function to fit the beam intensity profile, which was measured from the charge-coupled device (CCD) camera image of the YIG sample surface. Based on this measurement, the green probe laser had a beam diameter of 1.0 μm along the minor axis and 2.6 μm along the major axis, and the red heating laser had a beam diameter of 1.8 μm along the minor axis and 3.6 μm along the major axis. To confirm the obtained beam sizes, we conducted an additional beam size measurement by performing a micro-Raman scan on the top surface of a Si wafer with a sharp cleaved edge covered by an evaporated thin Au layer. With the laser beam incident normal to the Si surface, the obtained Raman Si peak intensity profile ($I(x)$) across the edge is shown in Fig. S2. The extracted value of the beam diameter for the green probe laser was 0.93 μm, which shows less than 10% difference from the other measurement result. If the spot sizes were smaller by 10%, the calculated phonon temperature increases by ~2 K at the heating laser power of 13.2 mW.

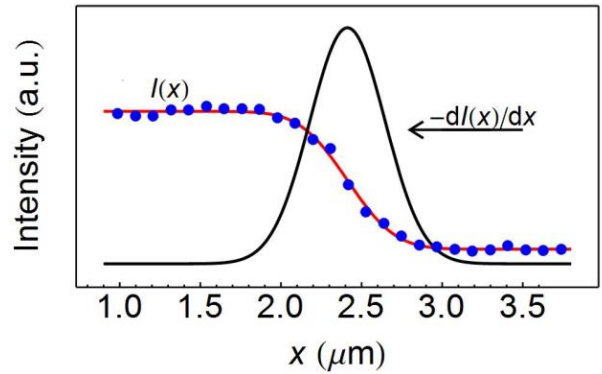

Fig. S2 Raman Si peak intensity profile measured across a cleaved Si edge.

**S5 Phonon BLS spectra under 0 K and 43 K stage temperature rise**

Here, we plot two phonon spectra at two stage heating conditions to show that there is indeed a frequency shift in the phonon spectra of approximately 0.14 GHz.

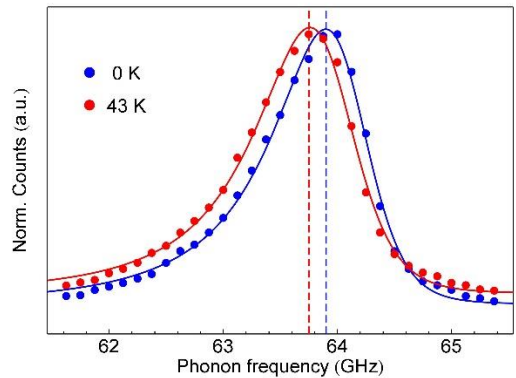

Fig. S3 BLS phonon spectra for the stage temperature rise of 0 K (blue) and 43 K (red) normalized to their maximum counts. The peak frequency is shifted by 0.14 GHz as quoted in the main text.

**S6 BLS spectra under laser heating condition**

We show the raw BLS spectra for the laser heating condition. Similar to the stage heating case, both magnon and phonon spectra shifts down in frequency with increasing laser power.

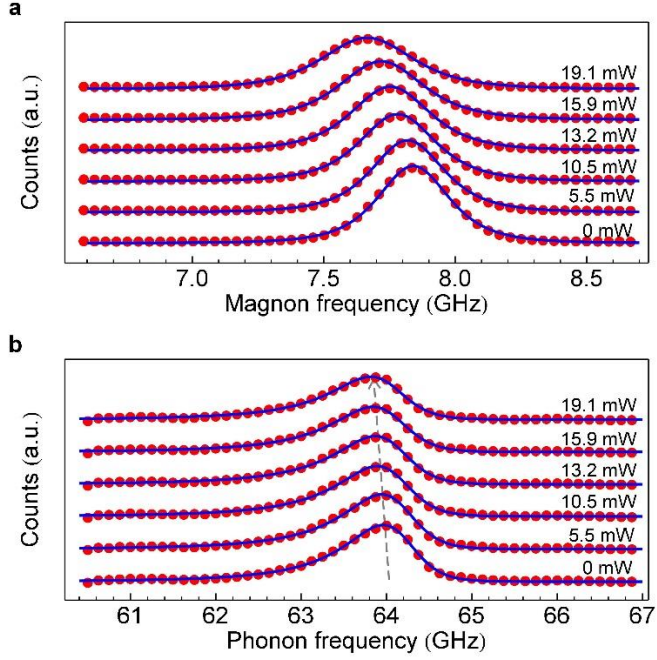

Fig. S4 BLS spectra for (**a**) magnons and (**b**) phonons, obtained with the YIG sample heated with the red heating laser at room temperature. Solid lines are fitting using (**a**) symmetric and (**b**) asymmetric squared Lorentzian functions. The arrow in (**b**) is drawn to show the small downward frequency shift of the phonon signal. The numbers inside the figures represent the red heating laser power.

**S7 Peak positions of BLS magnon and phonon spectra of YIG**

The peak position of the BLS magnon spectrum can be approximated by the following magnon dispersion relation [S5], which is derived for $\bm{q}$ perpendicular to the magnetization $\bm{M}$,

$$f_{\mathrm{m}} = \frac{\gamma}{2\pi}\sqrt{\begin{array}{c}(B + B_{\mathrm{a}} - \mu_0 NM + Dq^2) \\ \times (B + B_{\mathrm{a}} - \mu_0 NM + Dq^2 + \mu_0 M)\end{array}}, \quad (S1)$$

where $\gamma$ is the gyromagnetic ratio of $1.76 \times 10^{11}$ Hz T$^{-1}$, $\mu_0$ is the permeability of free space. $B_{\mathrm{a}}$ is the anisotropy field of 1 mT [S6], $N$ is the calculated demagnetization factor along the applied field direction of 0.08 [S7], $D$ is the exchange stiffness constant of $(5.4 \pm 0.1) \times 10^{-17}$ T m$^2$ [S8], and $M$ is the saturation magnetization. The peak frequency $f_{\mathrm{m}}$ corresponds to the frequency of the probed magnon mode with wave vector $q = 5.53 \times 10^7$ m$^{-1}$. To obtain $M$, the magnetic field dependent thermal magnon spectra were obtained with varying magnetic fields from 16.5 mT to 99.0 mT. Each spectrum was fitted with a Lorentzian function and the peak positions of the spectra were extracted and plotted in Fig. S5. By fitting Eq. (S1) to the H-field dependent peak frequency shift data, we obtain $M = (1.48 \pm 0.08) \times 10^5$ A/m at room temperature. This value agrees reasonably well with the literature [S8].

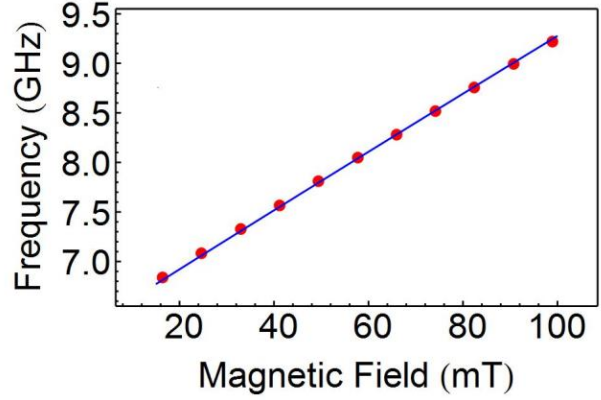

Fig. S5 BLS thermal magnon frequency as a function of external magnetic field.

The peak position of the phonon BLS spectrum is determined by the following equation for a linear phonon dispersion,

$$f_\mathrm{p} = \frac{v_\mathrm{q} q}{2\pi} = \frac{2 n v_\mathrm{q}}{\lambda}, \tag{S2}$$

where $v_\mathrm{q}$ is the phase velocity of the measured longitudinal acoustic phonon mode along the $q$ direction. The transverse acoustic phonon modes cannot be detected in our backscattering configuration [S6]. YIG can be treated as an elastically isotropic material because the elastic anisotropy factor 0.97 is very close to unity [S9]. We then calculate $v_\mathrm{q} = v_{[110]} = \sqrt{C_\mathrm{eff}/\rho}$, where $C_\mathrm{eff} = (C_{11} + C_{12} + 2C_{44})/2 = 270$ GPa and the density is $\rho = 5.17 \times 10^3$ kg/m$^3$ for YIG. The obtained phase velocity and phonon frequency are $v_\mathrm{q} = 7.2 \times 10^3$ m/s and $f_\mathrm{p} = 63.4$ GHz, respectively. This value agrees well with our observed peak position of 63.9 GHz in the phonon BLS spectrum at room temperature.

To accurately extract the peak position in the phonon BLS spectrum, we used an asymmetric squared Lorentzian function similar to that used in a previous work [S10],

$$L(f) = \left[\frac{d}{(f - f_0)^2 + \gamma(f, f_0)^2}\right]^2 + a,$$

where (S3)

$$\gamma(f, f_0) = \frac{2 f_0}{1 + e^{g(f - f_0)}},$$

where $d, f_0, \gamma,$ and $a$ are the fitting parameters, and $g$ is an additional fitting parameter for the asymmetry of data. A systematic fitting procedure using the Mathematica software package was applied to each phonon spectrum to extract the value of $f_0$. The standard error of $\pm\ 0.01$ GHz for the fitting parameter $f_0$ was obtained from the standard error in a Student's t-distribution of 8 and 6 spectra of red laser and stage heating measurements, respectively. This asymmetric function fits the data better than the symmetric Lorentzian function and allowed us to extract the peak position of spectrum more accurately. Nevertheless, we note that the difference in the peak positions determined by the symmetric and asymmetric functions is within the measurement uncertainty, as shown in Fig. S6.

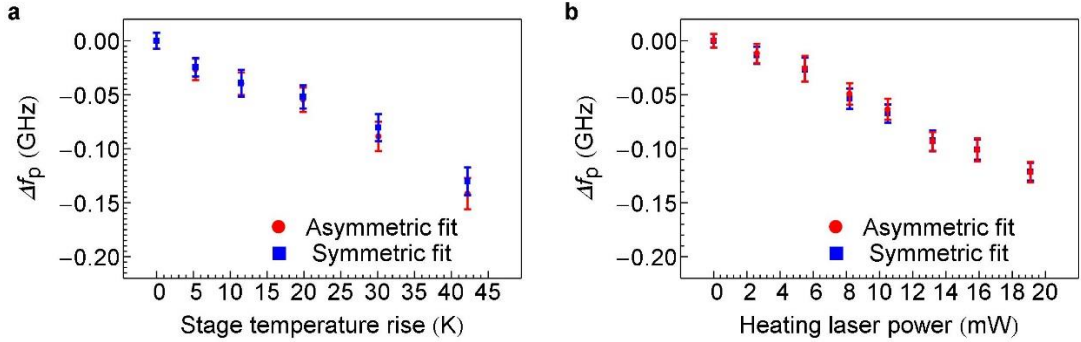

Fig. S6 Extracted peak positions in phonon BLS spectra with both symmetric and asymmetric fitting functions.

**S8 Measurement procedure and random error analysis**

The errors in the measured phonon or magnon frequencies are caused mainly by the relatively low signal-to-noise ratio of the BLS spectrum. To minimize the random errors, we took each phonon spectrum for 5 minutes and each magnon spectrum for 1 minute. Additionally, there was a long-term drift in the measured frequency, which can be caused by the thermal expansion of the optical cavities inside the BLS spectrometer during the long time required for the phonon measurements. There also could be a frequency drift (< 0.05 GHz/K) in our probing laser based on the specifications (Spectra physics Excelsior 532) due to room temperature fluctuations. To address the long term drift issue, we designed a specific measurement sequence in which we varied the heating laser power as 0, 2.6, 2.6, 0, 5.5, 5.5, 0, 8.2, 8.2, 0, 10.5, 10.5, 0, 13.2, 13.2, 0, 15.9, 15.9, 0, 19.1, 19.1, 0 mW, while the stage temperature was kept at room temperature. At each heating laser power, 8 spectra were collected. For the measurements with only stage heating, the stage temperature rise was varied in the sequence of 0, 5, 5, 0, 11, 11, 0, 20, 20, 0, 30, 30, 0, 43, 43, 0 K. This measurement sequence allowed us to monitor the frequency drift at zero heating, and thus improve the accuracy of the frequency shift at non-zero heating relative to the zero heating case. At each stage temperature rise, six spectra were collected. The frequencies obtained through this process are shown in Fig.

S7(a-b) and Fig. S8(a-b). The frequencies of the 8 and 6 spectra were averaged for the laser heating and stage heating measurements, respectively. The average frequencies are shown in Fig. S7(c-d) and Fig. S8(c-d). We then averaged the two sets of measurements taken at the same heating laser power or same stage temperature rise. Subsequently, the relative frequency change ($\Delta f$) was obtained by taking the difference between the frequency at non-zero laser (or stage) heating and that at the adjacent zero laser (or stage) heating. With a double-sided confidence of 95%, the random error in frequency was calculated from the standard error in a Student's t-distribution of 8 and 6 measurements for the laser and stage heating measurements, respectively. The random error in the phonon frequency was converted to a random error in the phonon temperature rise due to laser heating via the quadrature method of error propagation.

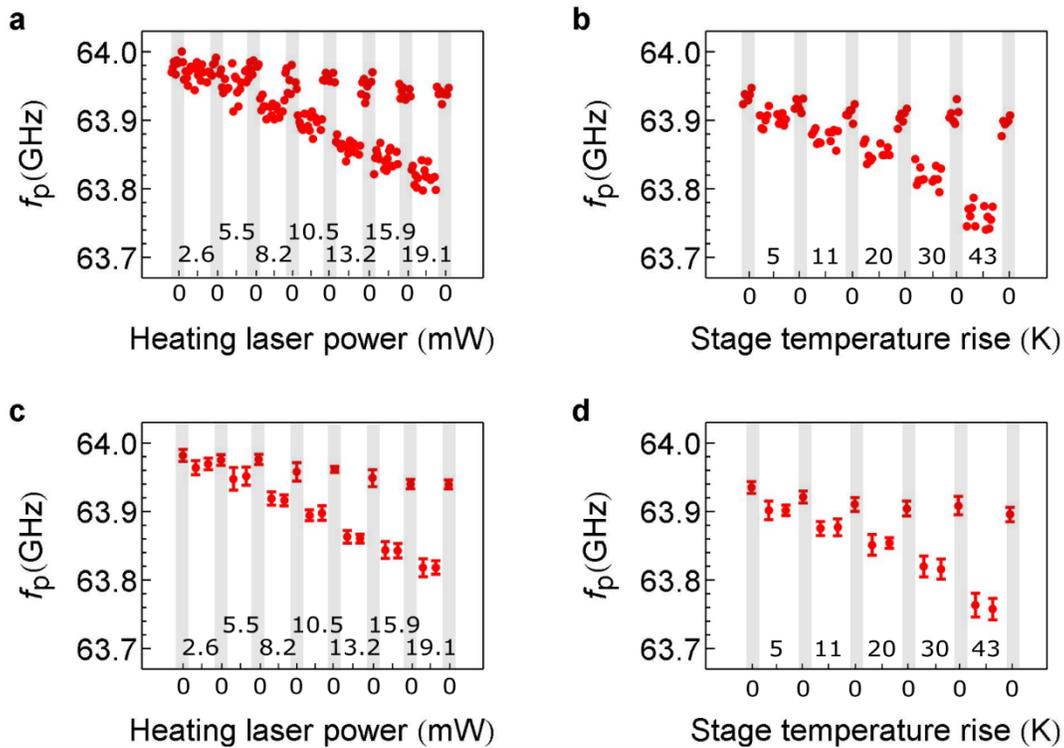

Fig. S7 **a.** Phonon frequency under laser heating. **b.** Phonon frequency under stage heating. **c.** Averaged phonon frequency under laser heating. **d.** Averaged phonon frequency under stage heating. Shaded area represents frequencies obtained without any heating by the red heating laser or the heater stage. For non-shaded areas in **a, b, c**, and **d**, the heating laser power or the stage temperature rises are shown.

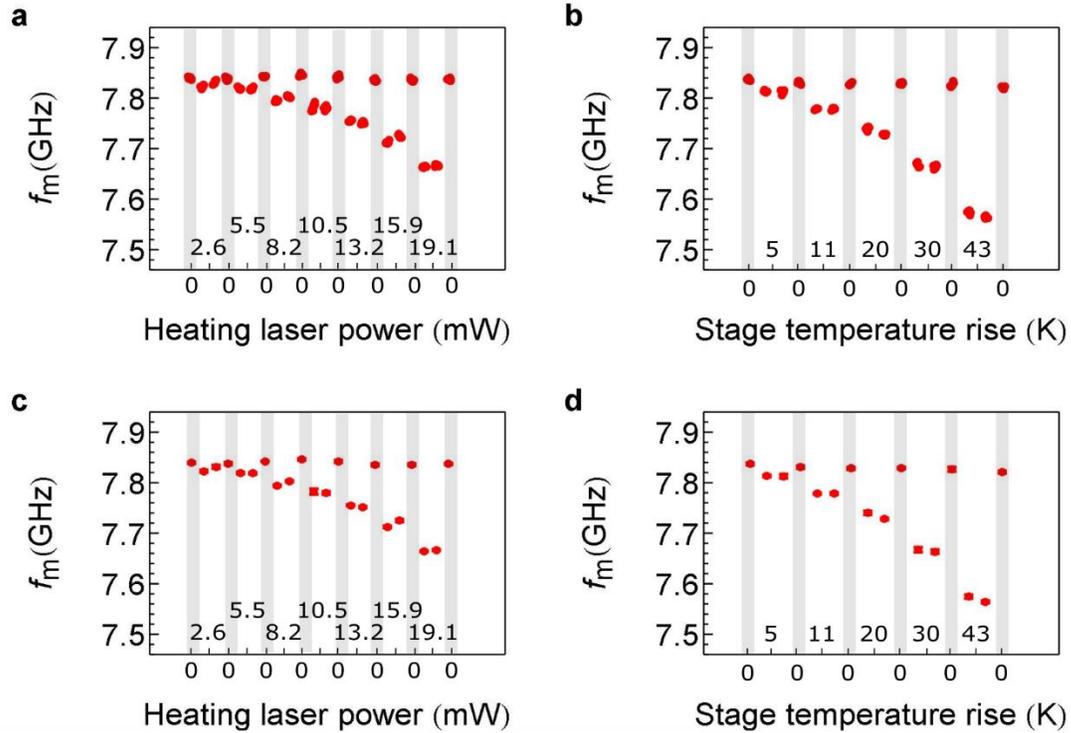

Fig. S8 **a.** Measured magnon frequency under laser heating. **b.** Measured magnon frequency under stage heating. **c.** Averaged magnon frequency under laser heating. **d.** Averaged magnon frequency under stage heating. Shaded area represents frequencies obtained without any heating by the red heating laser or the heater stage. For non-shaded areas in **a, b, c**, and **d**, the heating laser power or the stage temperature rises are shown.

**S9 Systematic error caused by strain effects on BLS phonon and magnon frequencies**

The induced stress ($\sigma$) could be different between the stage heating and the focused laser heating experiments. The difference ($\Delta\sigma$) in stress between the two experiments can potentially result in different peak shifts even when the lattice temperature rise ($\Delta T_\text{p}$) in the probe laser spot is the same. Hence, the effect of stress should be characterized and eliminated from the measurement results of the laser heating case to determine the phonon and magnon frequency shifts, $(\Delta f_\text{p})_\sigma$ and $(\Delta f_\text{m})_\sigma$, which would occur if the stress were the same as that for the stage heating condition at the same $\Delta T_\text{p}$.

For the phonon case,

$$(\Delta f_{\mathrm{p}})_{\sigma} = \left(\frac{\partial f_{\mathrm{p}}}{\partial T_{\mathrm{p}}}\right)_{\sigma} \Delta T_{\mathrm{p}} = \Delta f_{\mathrm{p}} - \left(\frac{\partial f_{\mathrm{p}}}{\partial \sigma}\right)_{T_{\mathrm{p}}} \Delta \sigma, \tag{S4}$$

where $\Delta f_{\mathrm{p}}$ is the measured phonon frequency shift. For the magnon case,

$$(\Delta f_{\mathrm{m}})_{\sigma} = \left(\frac{\partial f_{\mathrm{m}}}{\partial M}\right)_{\sigma} \Delta M = \Delta f_{\mathrm{m}} - \left(\frac{\partial f_{\mathrm{m}}}{\partial \sigma}\right)_{T_{\mathrm{p}}} \Delta \sigma, \tag{S5}$$

where $\Delta f_{\mathrm{m}}$ is the measured magnon frequency in the laser heating experiment. In this section, we will show calculations and measurements of the stress-correction terms, $\left(\frac{\partial f_{\mathrm{p}}}{\partial \sigma}\right)_{T_{\mathrm{p}}} \Delta\sigma$ and $\left(\frac{\partial f_{\mathrm{m}}}{\partial \sigma}\right)_{T_{\mathrm{p}}} \Delta\sigma$. In the backscattering configuration, the BLS phonon frequency is determined by Eq. (S2) and both the refractive index $n$ and phase velocity $v_{\mathrm{q}}$ depend on strain. The BLS magnon frequency is determined by Eq. (S1) and the probed magnon wave vector $q = 4\pi n/\lambda$ can depend on strain through the refractive index. The anisotropy field $B_{\mathrm{a}}$, the saturation magnetization $M$, and exchange stiffness constant $D$ can also depend on strain.

The strain and stress tensors were calculated with the COMSOL software package using the thermal stress module. For the stage heating case, we chose a boundary condition that allows free expansions along all directions for the top and side boundaries. The bottom surface of the YIG sample was attached with an epoxy to a ceramic heater plate, which has a lower thermal expansion coefficient than YIG. A similar type of tri-layer structure composed of die-epoxy-substrate was considered and the shear stress due to thermal expansion coefficient mismatch was calculated in a prior work [S11]. We modeled our tri-layer structure composed of YIG, epoxy, and heater according to the diagram shown in Fig. S9.

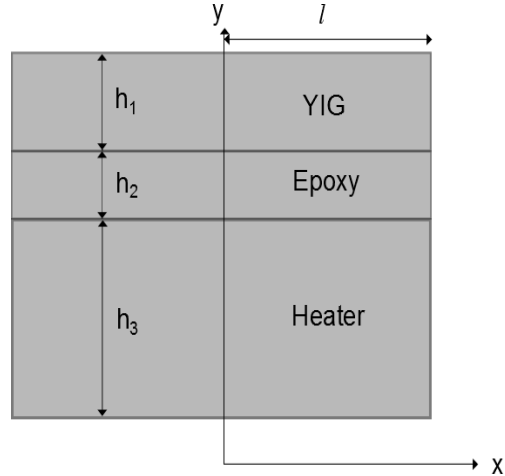

Fig. S9 Tri-layer structure used to simulate the YIG sample attached to a heater

The shear stress at the bottom of the YIG is a function of the radial coordinate $x$ and was calculated to be

$$\tau(x) = -\frac{k\Delta\alpha\Delta T \sinh kx}{\lambda \cosh kl}, \tag{S6}$$

where $\Delta\alpha = \alpha_1 - \alpha_3$ is the thermal expansion coefficient mismatch between the YIG (layer 1) and the heater (layer 3), $\Delta T$ is the temperature rise, 33 K, $l$ is the effective radius of the cylindrical tri-layer assembly, 1.6 mm, $k = \sqrt{\lambda/\kappa}$ is the assembly stiffness with axial compliance $\lambda$. The interfacial compliance $\kappa$ is expressed as

$$\lambda = \frac{1-\nu_1}{E_1 h_1} + \frac{1-\nu_3}{E_3 h_3} + \frac{h^2}{4D},$$

$$\kappa = \frac{h_1}{3G_1} + \frac{2h_2}{3G_2} + \frac{h_3}{3G_3},$$

where $h$ is the total thickness ($h = h_1 + h_2 + h_3$) and $D$ is the total flexural rigidity of the structure expressed as

$$D = \frac{E_1 h_1^3}{12(1-\nu_1^2)} + \frac{E_3 h_3^3}{12(1-\nu_3^2)}$$

The calculation was performed using the following values:

YIG: $E_1 = 200$ GPa, $G_1 = 77$ GPa, $\nu_1 = 0.29$, $h_1 = 0.4$ mm, $\alpha_1 = 8.3 \times 10^{-6}$/K
Epoxy resin (cured): $E_2 = 1.2$ GPa, $G_2 = 0.45$ GPa, $\nu_2 = 0.32$, $h_2 = 0.1$ mm
AlN heater: $E_3 = 308$ GPa, $G_3 = 119$ GPa, $\nu_3 = 0.29$, $h_3 = 2.5$ mm, $\alpha_3 = 4.5 \times 10^{-6}$/K

With the values above and a uniform temperature rise of 33 K caused by the heater, we obtained a compressive stress of $\tau(x) = -2.3 \times 10^6 \sinh(cx)$ Pa on the bottom of the YIG due to the larger thermal expansion of YIG than that of the heater, where $c = 310$ m$^{-1}$ and $x$ is the distance from the center in meters. For the focused laser heating case, we chose a boundary condition that allows free expansion for all surfaces because the temperature rise is highly localized and the thermal stress is negligible at regions far away from the laser spot. Since the stress and strain values are sensitive to the geometry, we chose a cylindrical geometry with a 1.6 mm radius and 0.4 mm thickness, which is close to our actual sample size of 4 mm × 2 mm × 0.4 mm. The mesh size was ~0.3 μm for the central 50 μm × 50 μm region and ~3 μm for the rest of the geometry.

    The simulations were carried out for a red laser heating power of 13.2 mW on top of a 8 mW green laser heating or stage heating of 33 K temperature rise on top of a 8 mW

green laser heating (see the section S12 of the Supplemental Material for details of simulation parameters). The measured phonon frequency shifts are the same for both cases so that the average phonon temperature rises in the laser spot are the same for these two simulations. Since a cylindrical coordinate basis was used in the simulation, the stress or strain tensors were transformed to a Cartesian coordinate basis through $\boldsymbol{\sigma}' = \boldsymbol{R_1}\boldsymbol{\sigma}\boldsymbol{R_1}^T$, where $\boldsymbol{\sigma}'$ and $\boldsymbol{\sigma}$ are the stress or strain tensors in Cartesian and cylindrical coordinates, respectively. $\boldsymbol{R_1}$ is the rotation matrix defined as $\boldsymbol{R_1} = \begin{pmatrix} \cos\phi & -\sin\phi & 0 \\ \sin\phi & \cos\phi & 0 \\ 0 & 0 & 1 \end{pmatrix}$. A weighted average of the stress or strain tensors over the probe laser beam were calculated in a similar manner as that described in the main text for the average temperature. The lasers were incident on the sample along the [110] out of plane direction, which is along the z axis. To express the tensors in a coordinate system where the x, y, and z axes are along the [100], [010], and [001] crystal axes, we performed a coordinate transformation $\boldsymbol{\sigma}'' = \boldsymbol{R_2}\boldsymbol{\sigma}'\boldsymbol{R_2}^T$, where $\boldsymbol{R_2} = \begin{pmatrix} 0 & 1/\sqrt{2} & 1/\sqrt{2} \\ 0 & -1/\sqrt{2} & 1/\sqrt{2} \\ 1 & 0 & 0 \end{pmatrix}$. The calculated weighted average stress and strain tensor in the probe laser spot for the laser heating and uniform stage heating cases are

$$\varepsilon_{\text{stage}} = \begin{pmatrix} 6.0 \times 10^{-4} & 2.0 \times 10^{-6} & 0 \\ 2.0 \times 10^{-6} & 6.0 \times 10^{-4} & 0 \\ 0 & 0 & 6.0 \times 10^{-4} \end{pmatrix},$$

$$\varepsilon_{\text{laser}} = \begin{pmatrix} 5.7 \times 10^{-4} & 8.6 \times 10^{-6} & 0 \\ 8.6 \times 10^{-6} & 5.7 \times 10^{-4} & 0 \\ 0 & 0 & 5.6 \times 10^{-4} \end{pmatrix},$$

$$\sigma_{\text{stage}} = \begin{pmatrix} -2.6 \times 10^{7} & 3.1 \times 10^{5} & 0 \\ 3.1 \times 10^{5} & -2.6 \times 10^{7} & 0 \\ 0 & 0 & -2.7 \times 10^{7} \end{pmatrix} \text{Pa},$$

$$\sigma_{\text{laser}} = \begin{pmatrix} -4.9 \times 10^{7} & 1.3 \times 10^{6} & 0 \\ 1.3 \times 10^{6} & -4.9 \times 10^{7} & 0 \\ 0 & 0 & -5.0 \times 10^{7} \end{pmatrix} \text{Pa}.$$

The values for the volume expansion ratio ($\Delta V/V$) were also calculated at each mesh point with the COMSOL software. The weighted averages of the volume expansion ratio over the probe laser spot size were used in the estimation of the magnon frequency shifts under thermal stress. Their values are 0.00162 and 0.00143 for the stage heating and the red laser heating condition, respectively. The smaller volume expansion for the red laser heating condition is expected because the volume expansion is suppressed due to the relatively cold surroundings.

## S7.1 Strain effect on refractive index

We first consider how the refractive index changes with non-zero strain. The change in the refractive index due to strain can be calculated from $\Delta n_i = -\frac{n_i^3}{2}\sum P_{ij}\varepsilon_j$ [S12], where the refractive index without strain is considered to be isotropic and the diagonal components are 2.34 for $\lambda = 532$ nm. $P_{ij}$ is the photo-elastic tensor and $\varepsilon_j$ is the strain tensor. Due to the symmetric property of the strain tensor, $\varepsilon_j$ can be expressed with six components instead of 9 using the Voigt notation, and a similar representation can be used for $\Delta n_i$. For a cubic crystal such as YIG, $P_{ij}$ has a simple form given by

$$P_{ij} = \begin{pmatrix} P_{11} & P_{12} & P_{12} & 0 & 0 & 0 \\ P_{12} & P_{11} & P_{12} & 0 & 0 & 0 \\ P_{12} & P_{12} & P_{11} & 0 & 0 & 0 \\ 0 & 0 & 0 & P_{44} & 0 & 0 \\ 0 & 0 & 0 & 0 & P_{44} & 0 \\ 0 & 0 & 0 & 0 & 0 & P_{44} \end{pmatrix},$$

where $P_{11} = 0.025$, $P_{12} = 0.073$, and $P_{44} = 0.041$ [S13].

For diagonal elements, we then obtain changes in the refractive index tensors of about $-0.00065$ and $-0.00062$ for the stage heating and the red laser heating condition, respectively, while off-diagonal elements are at least two orders smaller. The refractive index change alone gives rise to a phonon frequency change according to $\Delta f_p = \frac{\partial f_p}{\partial n}\Delta n$. The obtained values are $-0.0177$ GHz and $-0.0168$ GHz for the stage heating and the red laser heating condition, respectively. The difference is 0.0009 GHz, which is about an order of magnitude smaller than the random uncertainty in the measured phonon frequency, 0.01 GHz. The magnon frequency change caused by the refractive index change can be calculated as $\Delta f_m = \frac{\partial f_m}{\partial n}\Delta n$. The obtained values are $-0.0027$ GHz and $-0.0026$ GHz for the stage heating and the red laser heating condition, respectively. The difference is 0.0001 GHz, which is 30 times smaller than the random error in the measured magnon frequency, 0.003 GHz. Therefore, the systematic error caused by the refractive index change is negligible compared to the random error in both the magnon and phonon measurements by BLS.

## S7.2 Stress-induced phonon frequency shift

We now investigate the change in $f_p$ due to the change in the thermal stress. In a prior work on YIG [S14], the change in $f_p$ has been measured as a function of hydrostatic pressure in the range comparable to the laser heating induced stress. They obtained a value of $\frac{\Delta f_p}{f_p} = -\Delta\sigma \times 0.938 \times 10^{-11}$ Pa$^{-1}$ for the [110] longitudinal mode, where $\Delta\sigma$ is the applied hydrostatic pressure. The weighted averaged stresses in the laser spot are $-2.6 \times 10^7$ Pa and $-4.9 \times 10^7$ Pa for the stage heating and the red laser heating condition, respectively, where the negative sign indicates that the stress is compressive. The difference in stress can be treated approximately hydrostatic. With the difference in stress of $\Delta\sigma = -2.3 \times 10^7$ Pa and $f_\text{p} = 63.9$ GHz, we obtained $\Delta f_\text{p}$ of 13.8 MHz as a result of the thermal stress difference between the stage heating and the red laser heating condition. The difference of 0.0138 GHz is comparable to the random error in the measured phonon frequency, 0.01 GHz. In comparison, the measured phonon frequency change between 13.2 mW red laser heating and zero red laser heating is $-0.09 \pm 0.01$ GHz. Thus, the different stress conditions reduce the magnitude of the phonon frequency shift for the red laser heating condition by about 15% compared to the stage heating.

**S7.3 Strain-induced magnon frequency shift through magnetoelastic effect**

Magnetoelastic energy is affected by strain both via the change in the direction of magnetization with respect to the bond direction (the vector connecting two magnetic moments) and the change in the distance between magnetic moments. We consider the magnetoelastic energy density induced from a spatially uniform stress given by [S15]:

$$E_\sigma = A_1\left[\varepsilon_{11}\left(\alpha_1^2 - \frac{1}{3}\right) + \varepsilon_{22}\left(\alpha_2^2 - \frac{1}{3}\right) + \varepsilon_{33}\left(\alpha_3^2 - \frac{1}{3}\right)\right] \quad \text{(S7)}$$
$$+ A_2(\varepsilon_{12}\alpha_1\alpha_2 + \varepsilon_{23}\alpha_2\alpha_3 + \varepsilon_{31}\alpha_3\alpha_1),$$

where $\alpha_i (i = 1,2,3)$ are the directional cosines of the magnetization vector, $\varepsilon_{ij}(i,j = 1,2,3)$ are the strain tensors, and $A_1$ and $A_2$ are magneto elastic constants given as $A_1 = 3.48 \times 10^5$ Pa, $A_2 = 6.96 \times 10^5$ Pa [S16]. This additional energy leads to a reorientation of magnetization with respect to the bond direction between magnetic moments. This reorientation of magnetization can be described effectively by adding a strain induced field $B_\sigma$ to the external field. We can obtain $B_\sigma$ using the Smit – Suhl's uniform precession frequency formula given by

$$f = \frac{\gamma}{2\pi}\frac{1}{M\sin\theta}\left[\frac{\partial^2 E}{\partial \theta^2}\frac{\partial^2 E}{\partial \phi^2} - \left(\frac{\partial^2 E}{\partial\theta\partial\phi}\right)^2\right], \quad \text{(S8)}$$

where we use $E = -\boldsymbol{M}\cdot\boldsymbol{B}_\text{ext} + E_\sigma$. With $M = 1.48 \times 10^5$ A/m, $B_\text{ext} = 49.5$ mT, and the calculated strain tensors above, Eq. (S8) is used to obtain the effective field $B_\text{eff} = B_\text{ext} + B_\sigma$, where $B_\sigma$ is an additional field due to non-zero $E_\sigma$. We obtained $B_\sigma = -0.018$ mT and $B_\sigma = -0.080$ mT for the stage heating and the red laser heating condition, respectively. The magnon frequency shift due to $B_\sigma$ is $\Delta f_{\text{m},B_\sigma} = \frac{\partial f_\text{m}}{\partial B}\Delta B_\sigma$. As a result, we obtain $\Delta f_{\text{m},B_\sigma} = -0.0005$ GHz and $-0.0024$ GHz for the stage heating and the red laser heating condition, respectively.

Besides the reorientation of magnetization, the magnetoelastic energy depends on the distance between atomic magnetic moments. The pressure-dependent magnetization and anisotropy energy has previously been investigated [S17]. In terms of volume expansion, values of $dM/M = -1.57\, dV/V$ and $dB_a/B_a = -12\, dV/V$ were obtained. Using the previously obtained volume expansion ratio, we obtained $\Delta(\mu_0 M) = -0.47$ mT and $-0.42$ mT for the stage heating and the red laser heating condition, respectively. Similarly for $B_a$, we obtained $\Delta B_a = -0.019$ mT and $-0.017$ mT for the stage heating and the red laser heating condition, respectively. The magnon frequency change from $\Delta(\mu_0 M)$ is calculated by $\Delta f_{m,\mu_0 M} = \frac{\partial f_m}{\partial(\mu_0 M)}\Delta(\mu_0 M)$. We obtained $\Delta f_{m,\mu_0 M} = -0.0036$ GHz and $-0.0032$ GHz for the stage heating and the red laser heating condition, respectively. Similarly the magnon frequency change from $\Delta B_a$, is calculated by $\Delta f_{m,B_a} = \frac{\partial f_m}{\partial B_a}\Delta B_a$. We obtained $\Delta f_{m,B_a} = -0.00057$ GHz and $-0.00051$ GHz for the stage heating and the red laser heating condition, respectively.

The exchange stiffness parameter $D$ also depends on volume expansion. It has been reported [S18] that the exchange integral decreases with increasing volume as $dJ/J = -3.26\, dV/V$. Since the exchange parameter $D$ is defined as $D = Ja^2/\mu_B$, where $a$ is the lattice parameter and $\mu_B$ is the Bohr magneton. Then, the volume dependence of $D$ can be evaluated as

$$\Delta D = \Delta\left(\frac{Ja^2}{\mu_B}\right) = \frac{1}{\mu_B}(a^2\Delta J + J\Delta a^2) \approx \frac{a^2}{\mu_B}\left(\Delta J + \frac{2J\Delta V}{3V}\right)$$
$$= \frac{Ja^2\Delta V}{\mu_B V}\left(-3.26 + \frac{2}{3}\right) \approx -2.6\, D\frac{\Delta V}{V},$$

where the relation $3\Delta a/a = \Delta V/V$ was used to simplify the calculation. With $D = 5.4\times 10^{-17}$ T m$^2$, $q = 5.53\times 10^7$ m$^{-1}$, and the previously obtained volume expansion ratio, we obtained $\Delta B_{ex} = \Delta D q^2 = -0.69$ mT and $-0.61$ mT for the stage heating and the red laser heating condition, respectively. The magnon frequency change is calculated by $\Delta f_{m,B_{ex}} = \frac{\partial f_m}{\partial B_{ex}}\Delta B_{ex}$. We obtained $\Delta f_{m,B_{ex}} = -0.020$ GHz and $-0.018$ GHz for the stage heating and the red laser heating condition, respectively.

In summary, there are four sources for magnon frequency shift due to strain. The total frequency shift is then $\Delta f_m = \Delta f_{m,B_\sigma} + \Delta f_{m,B_{ex}} + \Delta f_{m,\mu_0 M} + \Delta f_{m,B_a} = -0.0246$ GHz and $-0.0241$ GHz for the stage heating and the red laser heating condition, respectively. The difference, 0.0005 GHz, is 6 times smaller than the random measurement uncertainty of 0.003 GHz.

**S7.4 Measurements of magnon frequency shift under hydrostatic pressure**

To verify the above theoretical calculation of the strain effect on the magnon frequency, we performed measurements of the magnon spectrum under a constant hydrostatic pressure that is comparable to the laser heating induced compressive stress, which is approximately isotropic based on the stress calculation results and much larger than the stress encountered in the stage heating measurement. The YIG sample was placed in a stainless steel pressure cell with an 8 mm c-axis sapphire sight window and connected to a pressurized argon cylinder. The pressure cell was placed in a magnetic field of 60 mT with the YIG sample mounted such that the field was aligned with the magnetic easy axis. Prior to mounting the sample, the magnetic field inside the pressure cell was calibrated to ensure that the stainless steel enclosure did not cause noticeable change in the magnetic field.

According to the simulations of the temperature-induced strain, the difference of weighted average stress between the stage heating and the red laser heating condition is approximately $2.3 \times 10^7$ Pa. This result was used to extrapolate an average stress for all other laser heating values, assuming a linear dependence of stress with laser heating. At each pressure, eight to fourteen spectra were acquired by a micro-BLS setup with two-minute acquisition time for each spectrum. A measurement at zero gauge pressure ($\sim 10^5$ Pa) was performed in between each high pressure measurement in the same way as was done for the laser power-dependent measurements. For each obtained spectrum, the magnon frequency was obtained as the average of the measured Stokes and anti-Stokes peak frequencies, so as to correct for the less than 0.01 GHz zero-point offset of the measured spectrum. The random error in the average magnon frequency of the eight to fourteen spectra was determined from the Student's t-distribution. The thick sapphire window was found to reduce the magnon signal by about one order of magnitude. Consequently, the random uncertainty in the measured magnon frequency of the YIG sample in the pressure cell increased up to 0.01 GHz, compared to about 0.003 GHz for the measurements outside the pressure cell. The difference in the measured magnon frequencies at different pressures are within the 0.01 GHz measurement uncertainty, except that the two zero pressure measurements immediately before and after the first $2.8 \times 10^7$ Pa measurement obtained about 0.01-0.02 GHz higher magnon frequencies than the other measurements, including the previous and successive zero pressure measurements. Based on these results, the magnon frequency change at the maximum pressure of $2.8 \times 10^7$ Pa is less than 0.02 GHz, which is less than 12% of the frequency change measured under the corresponding laser heating power of 19.1 mW.

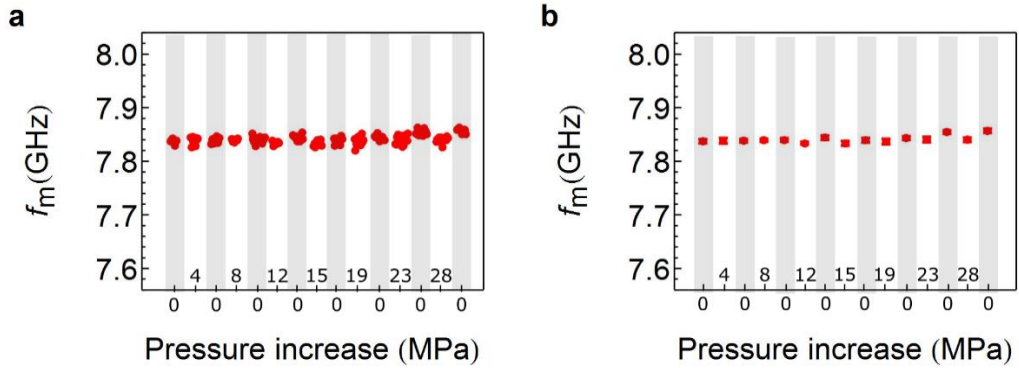

Fig. S10 **a**. Each data point is a two minute acquisition and each of the pressure measurements was repeated 8-14 times. Shaded areas represent frequencies obtained under zero gauge pressure. **b**. Average of 8-14 two minute acquisitions.

**S7.5 Summary of systematic error caused by strain**

So far we have considered how laser induced thermal stress would shift the magnon or phonon BLS frequencies. We have found that the change in refractive index is small and the corresponding changes in magnon and phonon BLS frequencies are negligible. In comparison, the phonon BLS frequency for the red laser heating case changes by about 15% due to the difference in stress from that in the stage heating case. Specifically, $\left(\frac{\partial f_p}{\partial \sigma}\right)_{T_p} \Delta\sigma$ is positive and about 15% of the measured frequency shift based on calculations (Fig. S11-a). Correction for this systematic error leads to an increase in the measured phonon temperature rise of 15% compared to the result without this correction (Fig. S11-b).

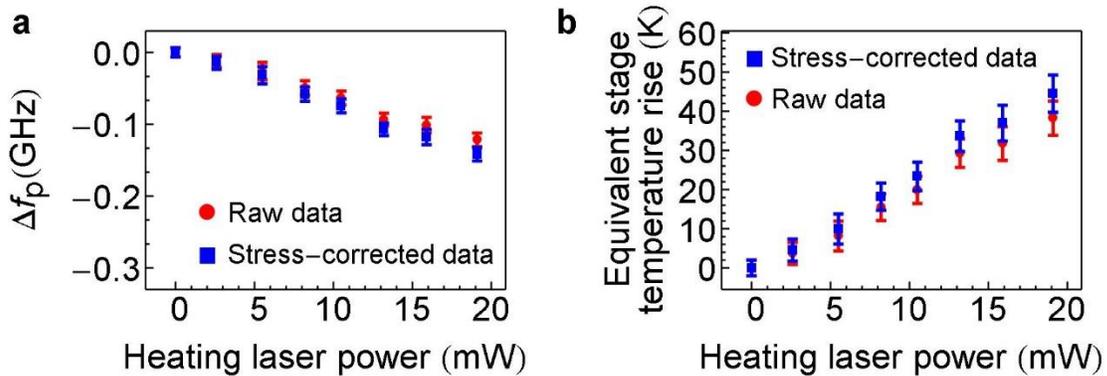

Fig. S11 **a.** Phonon frequency as a function of heating laser power $\Delta f_p$ (red disks) and data after the strain effect is removed $(\Delta f_p)_\sigma$ (blue rectangles). **b.** Equivalent stage temperature rise as a function of heating laser power.

For the magnon case, the theoretical analysis shows a negligible change in magnon frequency caused by strain. In addition, the pressure-dependent measurements verify that the laser-induced stress can decrease magnon frequency by less than 12% of the measured magnon frequency shift. Specifically, $\left(\frac{\partial f_\text{m}}{\partial \sigma}\right)_{T_\text{p}} \Delta \sigma$ is negative and less than 12% of the measured frequency shift (Fig. S12).

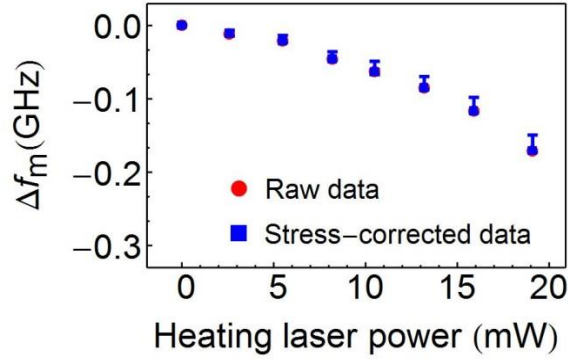

Fig. S12 Magnon frequency as a function of heating laser power (red disks) and data after the strain effect is taken account by increasing the length of error bars. (blue rectangles).

**S10 Magnon diffusion equation expressed in terms of magnon frequency**

In this section, we show a detailed process for rewriting the magnon diffusion equation (Eq. (1) in the main text) in terms of the magnon frequency deviation ($\delta f_\text{m}$). Equation (1) in the main text can be rewritten in terms of $\delta n = n - n_0$,

$$\nabla^2 n_0 + \nabla^2 \delta n = \frac{\delta n}{l_\text{r}^2}. \tag{S9}$$

The $\nabla^2 n_0$ term can be expanded as

$$\nabla^2 n_0 = \left(\frac{dn_0}{dT_\text{p}}\right)\nabla^2 T_\text{p} + \left(\frac{d^2 n_0}{dT_\text{p}^2}\right)|\nabla T_\text{p}|^2. \tag{S10}$$

$\delta n$ leads to a deviation in the local magnetization, $\delta M = -g\mu_B \delta n$, where $g$ and $\mu_B$ are the Landé $g$-factor and the Bohr magneton, respectively. Also $\delta M$ further gives rise to a deviation in the magnon peak frequency, $\delta f_m \approx \left(\frac{\partial f_m}{\partial M}\right)_H \delta M = -\left(\frac{\partial f_m}{\partial M}\right)_H g\mu_B \delta n$. Therefore,

$$\frac{dn_0}{dT_p} = -\frac{1}{g\mu_B}\left(\frac{\partial M}{\partial T_p}\right)_H,$$
$$\frac{d^2 n_0}{dT_p^2} = -\frac{1}{g\mu_B}\left(\frac{\partial^2 M}{\partial T_p^2}\right)_H. \tag{S11}$$

Using Eq. (S10) and (S11), Eq. (S9) can be rewritten as

$$\left(\frac{\partial M}{\partial T_p}\right)_H \nabla^2 T_p + \left(\frac{\partial^2 M}{\partial T_p^2}\right)_H |\nabla T_p|^2 + \nabla^2 \left[\frac{\delta f_m}{\left(\frac{\partial f_m}{\partial T_p}\right)_H}\left(\frac{\partial M}{\partial T_p}\right)_H\right]$$
$$= \frac{1}{l_r^2}\left[\frac{\delta f_m}{\left(\frac{\partial f_m}{\partial T_p}\right)_H}\left(\frac{\partial M}{\partial T_p}\right)_H\right], \tag{S12}$$

$M(T_p)$ of the YIG crystal used in this study was measured by using a vibration sample magnetometer (Fig. S13).

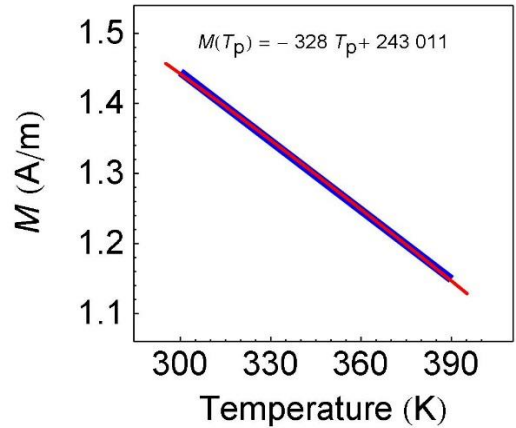

Fig. S13 The saturation magnetization of YIG as a function of temperature. The red line is the measured data. The blue solid line is a linear fit.

The measured $M(T_p)$ can be fit well to a linear relation, $M(T_p) = aT_p + b$ with $a = -328$ A m$^{-1}$ K$^{-1}$ and $b = -243{,}011$ A m$^{-1}$ in the temperature range relevant to the optical measurement.

Using this linear relation for $M(T_p)$, Eq. (S12) can be further simplified to the following, which is the same as Eq. (2) in the main text,

$$\nabla^2 T_p + \nabla^2 \delta\theta_m = \frac{\delta\theta_m}{l_r^2},$$

where

$$\delta\theta_m \equiv \frac{\delta f_m}{\left(\frac{\partial f_m}{\partial T_p}\right)_H}.$$

(S13)

To obtain $\left(\frac{\partial f_m}{\partial T_p}\right)_H$, we use the fit to the magnon data in Fig. 3a in the main text. The magnon frequency as a function of a stage temperature rise ($\Delta T_p^S$) is given by $f_m(\Delta T_p^S) = f_m(0) - 0.0038\,\Delta T_p^S, -0.000052\,(\Delta T_p^S)^2$. The absolute phonon temperature, $T_p$, can be expressed as $T_p = T_0 + \Delta T_p^S + \langle \Delta T_p^{S,G}\rangle$, where $T_0 = 302$ K is room temperature and $\langle\Delta T_p^{S,G}\rangle$ is the calculated weighted average of the phonon temperature rise in the probe laser spot due to the green probing laser at each stage temperature rise. $\langle\Delta T_p^{S,G}\rangle$ was numerically calculated by varying $\Delta T_p^S$ from 0 K to 50 K (Fig. S14). From the fit, we obtained $\langle\Delta T_p^{S,G}\rangle = \Delta T_p^{0,G} + 0.101\,\Delta T_p^S$, where $\Delta T_p^{0,G} = 34.5$ K. Therefore, $T_p$ can be expressed in terms of $\Delta T_p^S$ as $T_p = T_0 + \Delta T_p^{0,G} + 1.101\,\Delta T_p^S$. With this expression, we write $f_m$ in terms of $T_p$ and obtain $\left(\frac{\partial f_m}{\partial T_p}\right)_H = \left(\frac{\partial f_m}{\partial(\Delta T_p^S)}\right)_H \left(\frac{\partial(\Delta T_p^S)}{\partial T_p}\right)_H = 0.025 - 0.000086\,T_p$ in units of GHz / K.

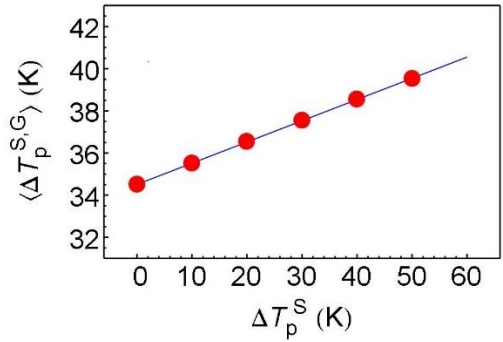

Fig. S14 Simulated weighted average of temperature rise due to the green laser heating as a function of stage temperature rise. The blue solid line is a fit to the data.

**S11 Derivation of energy transport equations**

The phonon temperature distribution is governed by the steady state energy equations for magnons and phonons, which can be obtained from the Boltzmann transport equation [S19-22],

$$\nabla \cdot (\kappa_m \nabla T_m) + g_{mp}(T_p - T_m) + Q_m = 0$$
$$\nabla \cdot (\kappa_p \nabla T_p) - g_{mp}(T_p - T_m) + Q_p = 0, \quad (S14)$$

where $\kappa_p$ and $\kappa_m$ are the phonon and magnon contributions to the temperature dependent total thermal conductivity ($\kappa = \kappa_p + \kappa_m$), $Q_p$ and $Q_m$ are the power densities of external heating absorbed by phonons and magnons, respectively, due to optical or electronic excitations associated with defects. $T_p$ and $T_m$ are the local phonon and magnon temperatures, respectively, and $g_{mp}$ is the magnon-phonon coupling or energy relaxation parameter [S19]. The magnon energy transport equation has been written for the case of uniform magnetic field [S22] and negligible magnon Peltier effect compared to heat diffusion [S21].

For the case of spatially uniform thermal conductivities, the phonon and magnon energy equations can be subtracted from each other to obtain the following equation that governs the local temperature difference, $\delta T \equiv T_p - T_m$, between phonons and magnons,

$$\nabla^2 \delta T - \frac{\delta T}{l_{mp}^2} + \frac{Q_p}{\kappa_p} - \frac{Q_m}{\kappa_m} = 0, \quad (S15)$$

where the magnon-phonon energy relaxation length is defined as $l_{mp} = [g_{mp}(1/\kappa_p + 1/\kappa_m)]^{-1/2}$ [S19]. We note that $\delta T$ is governed by $l_{mp}$, the characteristic magnon-phonon energy relaxation length scale, which is dominated by spin-preserving scattering processes, as opposed to $l_r$, which is determined by non-spin-preserving scattering processes.

The phonon and magnon energy equations can also be added to obtain Eq. (3) in the main text, which can be re-arranged as

$$\nabla \cdot (\kappa \nabla T_p) - \nabla \cdot (\kappa_m \nabla \delta T) + Q = 0, \quad (S16)$$

where $Q = Q_p + Q_m$ is the total power density of optical heating. It has been found that $\kappa_m$ is much smaller than $\kappa$ in YIG near room temperature [S3,23] and $|\nabla T_p|$ is larger than $|\nabla \delta T|$, so that

$$\nabla \cdot (\kappa \nabla T_\mathrm{p}) + Q \approx 0, \tag{S17}$$

where the temperature dependence of $\kappa(T_\mathrm{p})$ was extracted from a fit to the data from our measurement and the existing literature (see the section S2). Therefore, the phonon temperature profile in YIG mostly depends on the total heating power density and is insensitive to $\delta T$ because of a dominant $\kappa_\mathrm{p}$ compared to $\kappa_\mathrm{m}$ in YIG at room temperature.

**S12 Numerical simulation details**

To facilitate the simulation, we define an effective radius of the laser spot given by $\pi r_\mathrm{eff}^2 = \pi ab$, where $a$ and $b$ are the semi-major and semi-minor axis of the elliptical laser spot on the YIG surface (see the section S4 for the sizes of ellipses). The obtained $r_\mathrm{eff}$ were 0.8 μm and 1.3 μm for the probe laser and the heating laser, respectively. By employing an effective radius for the laser spot, the system becomes axially symmetric in the cylindrical coordinate, reducing the simulation to a two dimensional calculation. The simulation domain is defined by a cylinder of 200 μm in radius and 200 μm in thickness. Three different mesh sizes were used inside the cylinder. The mesh size is 0.1 μm for the central 10 μm × 10 μm regime, 0.4 μm in the next 50 μm × 50 μm regime and 5 μm for the rest of the cylinder.

We consider two cases in the simulation. In the first case, the stage temperature is kept at the room temperature (302 K), and the sample is heated by both the 13.2 mW red laser and the 8 mW green laser. In the second case, the stage temperature is kept at 335 K, and the sample is heated additionally by the 8 mW green laser only. These two cases were chosen because they yield the same measured phonon frequency shift. Correspondingly, the temperatures at the side and bottom of the bulk YIG are fixed at a temperature of 302 K and 335 K, respectively, for the two different cases. An adiabatic boundary condition is assumed for the top surface where the laser is incident. In addition, $\delta\theta_\mathrm{m}$ is set to be zero at the bottom and side boundaries of the simulation domain. The laser power density absorbed in YIG, $Q$, is calculated by taking derivative of the intensity profile $I(r,z)$ with respect to $z$. The intensity $I(r,z)$ is given by

$$I_\mathrm{i}(r,z) = \frac{2P_\mathrm{i}(1-R_\mathrm{i})}{\pi w_\mathrm{i}(z)^2} e^{-\frac{2r^2}{w_\mathrm{i}(z)^2} - \alpha_\mathrm{i} z}, \tag{S18}$$

where the subscript $i$ is either $G$ or $R$ for the green or red laser, respectively. The Gaussian beam divergence is described by $w_\mathrm{i}(z) = w_\mathrm{i}(0)\sqrt{1 + (z/z_\mathrm{i})^2}$ with $z_\mathrm{i} = \pi w_\mathrm{i}^2 n_\mathrm{i}/\lambda_i$. $R_\mathrm{G} = 0.16$ and $R_\mathrm{R} = 0.15$ are the reflectances [S24], $w_\mathrm{G} = 0.8$ μm and $w_\mathrm{R} = 1.3$ μm are the effective radii, $n_\mathrm{G} = 2.34$ and $n_\mathrm{R} = 2.27$ are the indexes of refraction [S24], $\alpha_\mathrm{G} = 1.5 \times 10^5$ m$^{-1}$ and $\alpha_\mathrm{R} = 5.0 \times 10^4$ m$^{-1}$ are the measured absorption coefficients for our YIG sample, which are close to the literature values [S24], $P_G = 8$ mW and $P_R = 13.2$ mW are the power of laser, $\lambda_G = 532$ nm and $\lambda_R = 660$ nm are the wavelengths. The power density $Q_i$ is then obtained from $Q_i = -\partial I_i(r,z)/\partial z$. With these parameters, Eq. (2) and Eq. (4) in the main text are solved numerically to obtain $\delta\theta_m(r,z)$ and $T_p(r,z)$.

In the following discussion, the superscripts $G$, $R$, and $S$ indicate heating by the 8 mW green laser, heating by the 13.2 mW red laser, and a stage temperature rise of 33 K, respectively. The phonon temperature profile for the first case with both laser heating and without the stage heating, $T_p^{R,G}(r,z)$, is expressed as $T_p^{R,G}(r,z) = T_0 + \Delta T_p^{R,G}(r,z)$, where $T_0 = 302$ K is the room temperature and $\Delta T_p^{R,G}(r,z)$ represents the change in the phonon temperature due to both lasers. Similarly, the temperature profile for the second case with both stage temperature rise 33 K and green laser heating but without the red laser heating, $T_p^{S,G}(r,z)$, is written as $T_p^{S,G}(r,z) = T_0 + \Delta T_p^S + \Delta T_p^{S,G}(r,z)$, where $\Delta T_p^S = 33$ K is the stage temperature rise and $\Delta T_p^{S,G}(r,z)$ is the corresponding change in the phonon temperature due to the green laser at this stage temperature. The difference in the phonon temperature rise between the two cases due to the red laser heating is obtained as $\Delta T_p^R(r,z) = \Delta T_p^{R,G}(r,z) - \Delta T_p^{S,G}(r,z)$. The weighted average value in the probe laser spot, $\langle \Delta T_p^R \rangle$, was 32 K, which is indeed close to the equivalent stage temperature rise 33 K caused by 13.2 mW red laser heating. The good agreement verifies the phonon temperature measurement by the BLS. In addition, we calculate the magnon frequency deviation profile between the two cases as given by $\delta f_m^R(r,z) = \delta f_m^{R,G}(r,z) - \delta f_m^{S,G}(r,z)$ (Fig. 4b in the main text), where $\delta f_m^{R,G} = \delta f_m^{S,G} = 0$ at the bottom of the simulation domain was used.

The weighted average values over the probe laser beam diameter were calculated with the $T_p(r,z)$ and $\delta f_m(r,z)$ using the following equation similar to that used in [S25],

$$\langle \xi \rangle = \frac{\int_0^\infty dz \int_0^\infty r dr\, \xi(r,z) Q_G(r,z)}{\int_0^\infty dz \int_0^\infty r dr\, Q_G(r,z)} \tag{S19}$$

where $Q_G(r,z) = -\partial I_G(r,z)/\partial z$ is the power density of the green probe laser and $\xi$ is either $T_p(r,z)$ or $\delta f_m(r,z)$ obtained from the simulations.

Finally, we show the detailed procedure of obtaining the magnon number density profiles (Fig. 4d). The magnon number deviation ($\delta n(r,z)$) from the local equilibrium value can be obtained by using $\delta n(r,z) = -\frac{\delta f_m(r,z)}{g\mu_B \left(\frac{\partial f_m}{\partial M}\right)_H} = -\frac{1}{g\mu_B} \left(\frac{\partial M}{\partial T_p}\right)_H \delta\theta_m(r,z)$, where $\delta\theta_m \equiv \frac{\delta f_m}{\left(\frac{\partial f_m}{\partial T_p}\right)_H}$ is the same as defined in the section S10. The local magnon number density can be obtained as $n(r,z) = \delta n(r,z) + n_0(T_p(r,z))$, where $n_0(T_p(r,z))$ is the equilibrium value given by the Bose-Einstein distribution at the local phonon temperature $T_p(r,z)$. The increase from the equilibrium value $n_0(T_0)$ at room temperature ($T_0$) is defined as $\Delta n(r,z) \equiv n(r,z) - n_0(T_0)$. This value is then normalized with $n_0(T_0)$ and plotted in Fig. 4d in the main text for both cases.